\documentclass[prd,superscriptaddress,amsfonts,amssymb,amsmath,showpacs,twocolumn]{revtex4-2}
\usepackage{bm}
\usepackage{amsfonts}
\usepackage{latexsym}
\usepackage[latin1]{inputenc}
\usepackage{graphicx}
\usepackage{amsmath}
\usepackage{palatino}
\usepackage{mathpazo}
\usepackage{textcomp}
\usepackage{float}
\usepackage{booktabs}
\usepackage{dcolumn}
\usepackage{booktabs}
\usepackage{multirow}
\usepackage{hyperref}
\hypersetup{colorlinks,citecolor=blue}
\usepackage{amsmath}
\usepackage{xcolor}
\usepackage{orcidlink}
\usepackage[caption=false]{subfig}
\usepackage{commath}
\def\jnl@style{\it}
\def\aaref@jnl#1{{\jnl@style#1}}

\def\aaref@jnl#1{{\jnl@style#1}}

\def\aj{\aaref@jnl{AJ}}                   
\def\apj{\aaref@jnl{ApJ}}                 
\def\apjl{\aaref@jnl{ApJ}}                
\def\apjs{\aaref@jnl{ApJS}}               
\def\apss{\aaref@jnl{Ap\&SS}}             
\def\aap{\aaref@jnl{A\&A}}                
\def\aapr{\aaref@jnl{A\&A~Rev.}}          
\def\aaps{\aaref@jnl{A\&AS}}              
\def\mnras{\aaref@jnl{Mon.~Not.~Roy.~Astron.~Soc.}}             
\def\prd{\aaref@jnl{Phys.~Rev.~D}}        
\def\prc{\aaref@jnl{Phys.~Rev.~C}}  
\def\prl{\aaref@jnl{Phys.~Rev.~Lett.}}    
\def\qjras{\aaref@jnl{QJRAS}}             
\def\skytel{\aaref@jnl{S\&T}}             
\def\ssr{\aaref@jnl{Space~Sci.~Rev.}}     
\def\zap{\aaref@jnl{ZAp}}                 
\def\nat{\aaref@jnl{Nature}}              
\def\aplett{\aaref@jnl{Astrophys.~Lett.}} 
\def\apspr{\aaref@jnl{Astrophys.~Space~Phys.~Res.}} 
\def\physrep{\aaref@jnl{Phys.~Rep.}}      
\def\physscr{\aaref@jnl{Phys.~Scr}}       
\def\commat{\aaref@jnl{Comm.~Math.~Phys.}}              
\def\science{\aaref@jnl{Science}}               
\def\cqg{\aaref@jnl{Classical Quant.~Grav.}}            
\def\jpcs{\aaref@jnl{JPCS}}                                     
\def\ijmpd{\aaref@jnl{Int.~J.~Mod.~Phys.~D}}                    
\def\grg{\aaref@jnl{Gen.~Relat.~Gravit.}}               
\def\rpp{\aaref@jnl{Rep.~Prog.~Phys.}}          
\def\npa{\aaref@jnl{Nucl.~Phys.~A}}        
\def\lrr{\aaref@jnl{Living Rev.~Rel.}}                   
\def\jcap{\aaref@jnl{J.~Cosmology Astropart.~Phys.}}    
\def\rmp{\aaref@jnl{Rev.~Mod.~Phys.}}   
\def\epjc{\aaref@jnl{Eur.~Phys.~J.~C}}

\allowdisplaybreaks[1]

\begin{document}

\color{black}       

\title{Exploring accelerated expansion in the universe: A study of $f(Q,T)$ gravity with parameterized EoS and cosmological constraints}

\author{M. Koussour\orcidlink{0000-0002-4188-0572}}
\email[Email: ]{pr.mouhssine@gmail.com}
\affiliation{Department of Physics, University of Hassan II Casablanca, Morocco.} 

\author{N. Myrzakulov\orcidlink{0000-0001-8691-9939}}
\email[Email: ]{nmyrzakulov@gmail.com}
\affiliation{L. N. Gumilyov Eurasian National University, Astana 010008,
Kazakhstan.}

\author{J. Rayimbaev\orcidlink{0000-0001-9293-1838}}
\email[Email: ]{javlon@astrin.uz}
\affiliation{New Uzbekistan University, Mustaqillik Ave. 54, Tashkent 100007, Uzbekistan.}
\affiliation{University of Tashkent for Applied Sciences, Gavhar Str. 1, Tashkent 100149, Uzbekistan.}
\affiliation{National University of Uzbekistan, Tashkent 100174, Uzbekistan.}

\author{A. Errehymy\orcidlink{0000-0002-0253-3578}}
\email[Email: ]{abdelghani.errehymy@gmail.com}
\affiliation{Astrophysics Research Centre, School of Mathematics, Statistics and Computer Science, University of KwaZulu-Natal, Private Bag X54001, Durban 4000, South Africa}

\author{Orhan Donmez\orcidlink{0000-0001-9017-2452}}
\email[Email: ]{orhan.donmez@aum.edu.kw}
\affiliation{College of Engineering and Technology, American University of the Middle East, Egaila 54200, Kuwait}

\date{\today}

\begin{abstract}
The study conducted in this research paper utilizes the $f(Q,T)$ gravity, where $Q$ represents non-metricity and $T$ represents the trace of the energy-momentum tensor, to investigate the accelerated expansion of the universe. To complete the study, an effective EoS with one parameter $\alpha$, is parameterized as $\omega _{eff}=-\frac{3}{\alpha (1+z)^{3}+3}$. The linear version of $f(Q,T)=-Q+\sigma T$ is also considered, where $\sigma$ is a constant. By constraining the model with six BAO points, 57 Hubble points, and 1048 Pantheon sample datasets, the parameters $\alpha$ and $\sigma$ are determined to best match the data. The cosmological parameters and energy conditions for the model are derived and examined. The results show that the model is in good agreement with observations, and can serve as a valuable starting point for analyzing FLRW models in the $f(Q,T)$ theory of gravity.
\end{abstract}

\maketitle

\section{Introduction}

\label{sec1}

It is thought that the expansion of our universe is accelerating. Dark Energy (DE), a mysterious type of energy with extremely high negative
pressure, is thought to be the cause of this. Various cosmological
observations have supported the existence of this sort of energy \cite{Riess,Perlmutter,D.J.,W.J., T.Koivisto,S.F.,R.R.,Z.Y.}. Its precise
nature, however, is still an unresolved issue that requires more research.
The cosmological constant is the earliest and most straightforward option
for DE, but it has two significant theoretical issues, including coincidence
and fine-tuning concerns \cite{weinberg/1989}. The coincidence problem arises from the fact that the energy density of DE and matter in the universe are of the same order of magnitude at the present epoch, despite evolving differently over cosmic history. The fine-tuning problem, on the other hand, relates to the extremely small but non-zero value of the cosmological constant required to match observations. The energy density associated with the cosmological constant is many orders of magnitude smaller than what would be expected based on quantum field theory calculations \cite{Steinhardt/1999}. To solve the DE
problem, two methods have been proposed: One involves studying the dynamics
of various DE models, and the other entails modifying the Hilbert--Einstein
action of General Relativity (GR), which results in altered theories of
gravity. The Equation of State (EoS) parameter $\omega _{DE}=\frac{p_{DE}}{%
\rho _{DE}}$, where $p_{DE}$ is really the pressure and $\rho _{DE}$ is the
energy density of DE, allows the DE models to be separated from the
cosmological constant (also known as $\Lambda $CDM model). The $\Lambda $CDM
model is supported by the EoS parameter value of $\omega _{0}=-1.084\pm 0.063$
and $\omega _{0}=-1.073\pm _{0.089}^{0.090}$ from observational sources,
Supernova data and WMAP, respectively \cite{Hinshaw,Planck2020}. The most well-known DE models are the
interactive DE models, such as the Chaplygin gas family, Braneworld,
holographic DE (HDE), agegraphic DE models, etc \cite{Kamenshchik,Sahni,Li,Cai}. Scalar field models, such as quintessence ($-1<\omega _{DE}<-\frac{%
1}{3}$), phantom ($\omega _{DE}<-1$), k-essence, tachyon, and quintom, are
also well-known \cite{Padmanabhan, Caldwell, Nojiri}. The quintessence and
phantom DE theories are also supported by the observed value of the EoS
parameter.

The $f(R)$ and $f(R,T)$ theories of gravity, where $R$ and $T$\ denote the
curvature scalar and the trace of the energy-momentum tensor, respectively,
are crucial in explaining the DE models among the different alternative
theories. The Refs. \cite{Copeland, Harko, Nojiri1} provide a thorough
overview of both modified theories of gravity and DE models. In addition to $f(R)$ and $f(R,T)$ theories, there has been significant research on other geometrically extended theories of gravity that aim to explain the accelerated cosmic expansion. For example, $f(R, L_m)$ gravity, where $L_m$ is the matter Lagrangian density, has been proposed as an alternative to GR \cite{Harko/2010,THK-2,THK-3,V.F.-2}. Similarly, $f(G)$ gravity, where $G$ is the Gauss-Bonnet invariant \cite{fG1}, and $f(R,G)$ gravity, which combines modifications to the gravitational action involving both $R$ and $G$ \cite{Laurentis/2015,Gomez/2012}, have also been proposed as alternative theories of gravity that could explain the accelerated cosmic expansion. All of the aforementioned extensions of GR share the fundamental Riemannian geometry
that forms the basis of such classical concepts, especially GR \cite{Riemann}. Due to the inconsistency of these concepts at certain scales, it follows
logically that if the fundamental geometry could be replaced with a much
more universal geometric framework, we might be able to eliminate some of
the contradictions (such as the dark matter (DM) and DE that have
dogged these classical concepts over the years. Weyl made such an innovative
effort, where the geometric unification of gravity and magnetism was the
primary goal \cite{Weyl}. We are aware that the Levi-Civita connection,
which is the fundamental method for comparing the lengths of vectors in
Riemannian geometry, is compatible with the metric. The mechanism utilized
in Weyl's theory has two connections, one of which contains the vector's
length information and the other of which determines the vector's direction
throughout parallel transport. The electromagnetic potential is physically
linked to the length connection. The non-zero covariant divergence of the
metric tensor i.e. $\nabla _{\gamma }g_{\mu \nu }\neq 0$, which is a trait
that leads to a additional geometric quantity called the non-metricity $Q$,
is the most remarkable aspect of the theory. There are two different
representations of GR that we can find in the literature: the first one has $%
R\neq 0$, $T=0$ (here $T$ is the torsion), and $Q=0$ (curvature
representation), while the second one has $R=0$, $T\neq 0$, and $Q=0$
(teleparallel representation) \cite{Moller, Pellegrini, Hayashi}. Therefore,
the non-metricity $Q$ disappears in each of these representations.
Geometrically, $Q$ depicts the variation in a vector's length in parallel
transport. Now, a non-vanishing non-metricity $Q$ was thought of as the
fundamental geometrical variable accountable for all varieties of
gravitational interactions in a third equivalent representation of GR. The
symmetric teleparallel gravity (STG) is the name given to this theory \cite%
{Jimenez1}. In \cite{Jimenez2, Lazkoz}, the cosmology of $f(Q)$ gravity and
its observational constraints were examined. There have been a number of
publications in the STG framework during the past few decades \cite{Frusciante, MK1, MK2, MK3,MK4,MK5,MK6,MK7,MK8}.

According to the authors of Ref. \cite{Harko1}, the non-minimal coupling
between the matter Lagrangian $L_{m}$ and the non-metricity scalar $Q$
allows for an extension of the $f(Q)$ theory of gravity. As would be
predicted, the non-minimal coupling between the geometry and matter sectors
causes the energy-momentum tensor to not be conserved and causes an
additional force to appear in the geodesic equation of motion. Another
generalization of the $f(Q)$\ theory, known as the $f(Q,T)$ theory, was
proposed by Xu et al. \cite{Xu}, where the gravity Lagrangian is essentially
an arbitrary function of the non-metricity scalar $Q$ and the trace of the
energy-momentum tensor $T$. The cosmic evolution was investigated while the
field equations were being developed. It was discovered that the theory
indicated an accelerated expansion of the universe culminating in a
de-Sitter type evolution in all situations taken into consideration. In $%
f(Q,T)$ gravity, the late-time cosmology in the hybrid expansion law \cite%
{Pati} and the bouncing scenarios \cite{Ag} are examined. Also, baryogenesis
in $f(Q,T)$ gravity was studied by Bhattacharjee et al. \cite{Bhattacharjee}.

The work is structured as follows: in Sec. \ref{sec2}, the action of $f(Q,T)$
gravity and fundamental field equations are provided. In Sec. \ref{sec3}, we
discussed the parametrization of the effective EoS parameter and the energy
conditions. We have discussed and analyzed our model in $f(Q,T)$ gravity
using some observational datasets in Sec. \ref{sec3}. In Sec. \ref{sec4}%
, the findings and conclusions are discussed.

\section{$f(Q,T)$ gravity theory}

\label{sec2}

The action used to determine $f(Q,T)$ gravity is defined as \cite{Xu},%
\begin{equation}
S=\int \sqrt{-g}d^{4}x\left( \frac{1}{16\pi }f(Q,T)+L_{m}\right) .  \label{1}
\end{equation}

Here, $f(Q,T)$ is an arbitrary function that correlates the trace of the
energy-momentum tensor $T$ to its non-metricity $Q$. Additionally, $L_{m}$
stands for the matter Lagrangian, and $g=det(g_{\mu \nu })$. The
non-metricity $Q$ is also described as \cite{Jimenez1},%
\begin{equation}
Q\equiv -g^{\mu \nu }(L_{\,\,\,\alpha \mu }^{\beta }L_{\,\,\,\nu \beta
}^{\alpha }-L_{\,\,\,\alpha \beta }^{\beta }L_{\,\,\,\mu \nu }^{\alpha }),
\label{2}
\end{equation}%
where $L_{\,\,\,\alpha \gamma }^{\beta }$ is the abbreviation for the
disformation tensor, 
\begin{eqnarray}
L_{\,\,\,\alpha \gamma }^{\beta } &=&-\frac{1}{2}g^{\beta \eta }(\nabla _{\gamma
}g_{\alpha \eta }+\nabla _{\alpha }g_{\eta \gamma }-\nabla _{\eta }g_{\alpha
\gamma }).  \label{3} \\
&=&\frac{1}{2}g^{\beta \eta }\left( Q_{\gamma \alpha \eta }+Q_{\alpha \eta
\gamma }-Q_{\eta \alpha \gamma }\right) , \\
&=&{L^{\beta }}_{\gamma \alpha }.
\end{eqnarray}

The tensor of non-metricity is denoted by, 
\begin{equation}
Q_{\gamma \mu \nu }=-\nabla _{\gamma }g_{\mu \nu }=-\partial _{\gamma
}g_{\mu \nu }+g_{\nu \sigma }\widetilde{\Gamma }{^{\sigma }}_{\mu \gamma
}+g_{\sigma \mu }\widetilde{\Gamma }{^{\sigma }}_{\nu \gamma },  \label{4}
\end{equation}%
where $\widetilde{\Gamma }{^{\sigma }}_{\mu \gamma }$ is the Weyl--Cartan
connection \cite{Xu}, and the trace of the non-metricity tensor being
provided as, 
\begin{equation}
Q_{\beta }=g^{\mu \nu }Q_{\beta \mu \nu },\qquad \widetilde{Q}_{\beta
}=g^{\mu \nu }Q_{\mu \beta \nu }.  \label{5}
\end{equation}

A superpotential or the non-metricity conjugate can also be defined as, 
\begin{eqnarray}
\hspace{-0.5cm} &&P_{\ \ \mu \nu }^{\beta }\equiv \frac{1}{4}\bigg[-Q_{\ \
\mu \nu }^{\beta }+2Q_{\left( \mu \ \ \ \nu \right) }^{\ \ \ \beta
}+Q^{\beta }g_{\mu \nu }-\widetilde{Q}^{\beta }g_{\mu \nu }  \notag \\
\hspace{-0.5cm} &&-\delta _{\ \ (\mu }^{\beta }Q_{\nu )}\bigg]=-\frac{1}{2}%
L_{\ \ \mu \nu }^{\beta }+\frac{1}{4}\left( Q^{\beta }-\widetilde{Q}^{\beta
}\right) g_{\mu \nu }-\frac{1}{4}\delta _{\ \ (\mu }^{\beta }Q_{\nu )}.
\label{6}
\end{eqnarray}%
\newline
expressing the scalar of non-metricity as \cite{Jimenez1},%
\begin{eqnarray}
&&Q=-Q_{\beta \mu \nu }P^{\beta \mu \nu }=-\frac{1}{4}\big(-Q^{\beta \nu
\rho }Q_{\beta \nu \rho }+2Q^{\beta \nu \rho }Q_{\rho \beta \nu }  \notag \\
&&-2Q^{\rho }\tilde{Q}_{\rho }+Q^{\rho }Q_{\rho }\big).
\end{eqnarray}

As a result, by equating the variation of action in Eq. \eqref{1} with regard to
the metric tensor to zero i.e. $\delta S=0$, we obtain the following field
equations: 
\begin{multline}
-\frac{2}{\sqrt{-g}}\nabla _{\beta }(f_{Q}\sqrt{-g}P_{\,\,\,\,\mu \nu
}^{\beta }-\frac{1}{2}fg_{\mu \nu }+f_{T}(T_{\mu \nu }+\Theta _{\mu \nu })
\label{11} \\
-f_{Q}(P_{\mu \beta \alpha }Q_{\nu }^{\,\,\,\beta \alpha }-2Q_{\,\,\,\mu
}^{\beta \alpha }P_{\beta \alpha \nu })=8\pi T_{\mu \nu }.
\end{multline}

Here, $f_{Q}=\dfrac{df}{dQ}$, $f_{T}=\dfrac{df}{dT}$ and $T_{\mu \nu }$ is
the energy-momentum tensor for the fluid of the ideal type, as described by%
\begin{equation}
T_{\mu \nu }=-\frac{2}{\sqrt{-g}}\dfrac{\delta (\sqrt{-g}L_{m})}{\delta
g^{\mu \nu }}  \label{8}
\end{equation}%
and 
\begin{equation}
\Theta _{\mu \nu }=g^{\alpha \beta }\frac{\delta T_{\alpha \beta }}{\delta
g^{\mu \nu }}.  \label{9}
\end{equation}

Moreover, the variation of energy-momentum tensor with respect to the metric
tensor is, 
\begin{equation}
\frac{\delta \,g^{\,\mu \nu }\,T_{\,\mu \nu }}{\delta \,g^{\,\alpha \,\beta }%
}=T_{\,\alpha \beta }+\Theta _{\,\alpha \,\beta }\,.  \label{10}
\end{equation}

Now, suppose the universe can be represented by the homogeneous, isotropic,
and spatially flat FLRW metric, 
\begin{equation}
ds^{2}=-dt^{2}+a^{2}(t)\left[ dx^{2}+dy^{2}+dz^{2}\right] ,  \label{12}
\end{equation}%
where $a(t)$ is the scale factor of the universe used to estimate the rate
of cosmic expansion at a time $t$. Further, it is presumed that the known
universe matter is made up of a perfect fluid, for which the energy-momentum
tensor, $T_{\,\,\,\nu }^{\mu }=diag(-\rho ,p,p,p)$. Moreover, the
non-metricity scalar $Q$ for this type of metric is derived and given as $%
Q=6H^{2}$, where $H$ is the Hubble parameter.

The generalized Friedmann equations are given below by using the metric (\ref{12}) and the field equation \eqref{11},
\begin{equation}
8\pi \rho =\frac{f}{2}-6FH^{2}-\frac{2\widetilde{G}}{1+\widetilde{G}}(\dot{F}%
H+F\dot{H}),  \label{13}
\end{equation}%
\begin{equation}
8\pi p=-\frac{f}{2}+6FH^{2}+2(\dot{F}H+F\dot{H}),  \label{14}
\end{equation}%
where, the dot ($\cdot $) denotes a derivative with respect to time, while
the symbols $F=f_{Q}$, and $8\pi \widetilde{G}=f_{T}$, respectively, signify
differentiation with respect to $Q$, and $T$.

Using the two Eqs. \eqref{13} and \eqref{14} mentioned above , we can
construct the equations similar to the form of standard GR, 
\begin{equation}
3H^{2}=8\pi \rho _{eff}=\frac{f}{4F}-\frac{4\pi }{F}\left[ (1+\widetilde{G}%
)\rho +\widetilde{G}p\right] ,  \label{15}
\end{equation}%
and 
\begin{multline}
2\dot{H}+3H^{2}=-8\pi p_{eff}=\frac{f}{4F}-\dfrac{2\dot{F}H}{F}+  \label{16}
\\
\frac{4\pi }{F}\left[ (1+\widetilde{G})\rho +(2+\widetilde{G})p\right] ,
\end{multline}%
where the terms $\rho _{eff}$, and $p_{eff}$ refer to the effective pressure
and density, respectively.

\section{Parametrization of the effective EoS parameter}

\label{sec3}

The system of field equations, discussed as Eqs. \eqref{13} and \eqref{14}, contains only two independent equations with four unknowns: $f(Q,T)$, $\rho$, $p$, and $H$. Therefore, two additional constraint equations are required to fully solve the system and study the temporal evolution of the energy density and pressure. The use of these equations is supported by several justifications in the literature, such as the model-independent way approach to studying DE models \cite{Pacif1,Pacif2}. This approach typically involves a parametrization of kinematic variables, such as the Hubble parameter, deceleration parameter, EoS parameter, and jerk parameter, and provides the necessary supplementary equation. The primary advantage of this method is that it allows for the examination of cosmological models using observational data. The relationship between the scale
factor $a(t)$ and the redshift $z$ is known to be represented by $\frac{a_{0}%
}{a}=1+z$, where $a_{0}$ is the present value of scale factor ($a_{0}=a(0)=1$%
). From the relationship mentioned above, we may conclude $\frac{d}{dt}=%
\frac{dz}{dt}\frac{d}{dz}=-(1+z)H(z)\frac{d}{dz}$. Thus, the Hubble
parameter can be expressed mathematically as, 
\begin{equation}
\dot{H}=-(1+z)H(z)\frac{dH}{dz}.  \label{19}
\end{equation}

In the context of this investigation, we will employ the following functional expression: $f(Q,T)=\chi Q+\sigma T$, where $\chi$ and $\sigma $ are model parameters. This model was first proposed by Xu et al. \cite{Xu} to describe an exponentially expanding universe, where the energy density $\rho$ scales as $\exp(-H_0 t)$. It has since been constrained by observational data related to the Hubble parameter, as detailed in \cite{Arora/2020}. Loo et al. \cite{Loo/2023} utilized this model to study Bianchi type-I cosmology, incorporating observational datasets such as Type Ia supernovae and the Hubble parameter. In addition, \cite{Tayde/2022} explored wormhole solutions within this model, considering various EoS relations. Hence, we have $F=f_{Q}=\chi$ and $8\pi \widetilde{G}=f_{T}=\sigma$. In this scenario, setting $\chi=-1$ and $\sigma=0$ leads to the well-motivated case of General Relativity (GR). In our study, we choose $\chi = -1$ to facilitate the derivation of solutions corresponding to GR. In addition, it's worth noting that these solutions remain independent of the parameter $\chi$. Next, the Friedmann
equations \eqref{13} and \eqref{14} for this particular $f(Q,T)$ model give
the following expressions for the energy density and pressure in terms of
redshift,%
\begin{equation}
\rho =\frac{3 H^2+\dot{H}}{8 \pi +2 \sigma}-\frac{\dot{H}}{8 \pi +\sigma },
\end{equation}%
and%
\begin{equation}
p=-\frac{3 H^2+\dot{H}}{8 \pi +2 \sigma }-\frac{\dot{H}}{8 \pi +\sigma }.
\end{equation}

The EoS parameter $\omega=\frac{p}{\rho }$ was determined as the
effective or total EoS parameter, 
\begin{equation}
\omega=\frac{3 H^2 ( 8 \pi +\sigma)+\dot{H} (16 \pi +3 \sigma)}{\sigma \dot{H} -3 H^2 (8 \pi +\sigma)}.  \label{17}
\end{equation}

Now, a single assumption is necessary to solve the system of equations, specifically Eqs. \eqref{13} and \eqref{14}. Recent studies have investigated various parametrizations of the EoS. The Chevallier-Polarski-Linder (CPL) parametrization, which is based on a simple Taylor expansion of the EoS with respect to the scale factor, is the most widely used parametrization. It is represented by $\omega=\omega_{0}+\omega_{1}\frac{z}{1+z}$ \cite{CPL1,CPL2}. Other popular parametrizations include the Jassal-Bagla-Padmanabhan (JBP) parametrization, which permits the transition from a DE-dominated universe to a matter-dominated universe, and is expressed as $\omega=\omega_{0}+\omega_{1}\frac{z}{(1+z)^2}$ \cite{JBP}, and the Ma-Zhang (MZ) parametrization, which is based on a logarithmic and oscillating form of the EoS. Specifically, the MZ parametrization has two forms: $\omega=\omega_{0}+\omega_{1}(\frac{\ln(2+z)}{1+z}-\ln2)$ and $\omega=\omega_{0}+\omega_{1}(\frac{\sin(1+z)}{1+z}-\sin(2))$ \cite{MZ}. The paper under consideration examines the following parametric form of the effective (or total) EoS as a function of redshift $z$ \cite{Mukherjee}, 
\begin{equation}
\omega _{eff}=-\frac{3}{\alpha (1+z)^{3}+3},  \label{20}
\end{equation}%
where $\alpha $ is a constant parameter. This specific form of $\omega_{eff}$ is motivated by many studies in the literature that utilize single-parameter EoS, which are commonly used in cosmology and astrophysics for their simplicity and effectiveness in capturing key aspects of the system \cite{Mukherjee,Gong/2005,Yang/2019}. These models have proven successful in describing a variety of phenomena, highlighting their importance in theoretical and observational studies of the universe. Also, we chose this form because it shows interesting behavior during the early stages of the universe's evolution. At large values of $z$, $\omega_{eff}$ is close to zero, resembling the EoS of non-pressurized fluid like DM. As the universe evolves, $\omega_{eff}$ becomes increasingly negative, indicating negative pressure. Setting $\alpha$ to zero results in $\omega_{eff}$ being -1, equivalent to a cosmological constant. For $\alpha$ less than -1, the model shows phantom behavior, with $\omega_{eff}$ dropping below -1. Unlike traditional DE models that are limited to the early stages and later dominate over DM, this model behaves like pressureless DM in its early stages, when $\omega_{eff}$ approaches zero \cite{Mukherjee}.

The differential equation for $H\left( z\right) $ is given by Eqs. (\ref{19}%
) and (\ref{17}) with the assumed ansatz of $\omega _{eff}$ as shown in Eq. (%
\ref{20}),

\begin{widetext}
\begin{equation}
\frac{-3(8\pi +\sigma )H^{2}+(16\pi +3\sigma )(1+z)H(z)\frac{dH}{dz}}{\sigma
(1+z)H(z)\frac{dH}{dz}+3(8\pi +\sigma )H^{2}}=-\frac{3}{\alpha (1+z)^{3}+3}
\end{equation}

Thus, the solution derived for the Hubble parameter $H\left( z\right) $ as
a function of redshift $z$\ is,
\begin{equation}
H\left( z\right) =H_{0}\left[ \frac{12\sigma +3\alpha \sigma (z+1)^{3}+16\pi
\alpha (z+1)^{3}+48\pi }{3(\alpha +4)\sigma +16\pi (\alpha +3)}\right] ^{%
\frac{\sigma +8\pi }{3\sigma +16\pi }},  \label{21}
\end{equation}%
\end{widetext}
where $H_{0}$ represents the present value (i.e. at $z=0$) of the Hubble
parameter. Specifically, in the case where $\sigma=0$, the solution simplifies to $f(Q,T) =-Q$. This directly corresponds to the $\Lambda$ CDM model. Consequently, the Hubble parameter equation $H\left( z\right) $ can be expressed as
\begin{equation}
    H\left( z\right) =H_{0}\left[\Omega _{m0}\left( 1+z\right) ^{3}+\Omega _{\Lambda }\right] ^{\frac{1%
}{2}}
\end{equation}
 where $\Omega _{m0}=\frac{\alpha}{3+\alpha}$ and $\Omega _{\Lambda
}=( 1-\Omega _{m0}) =\frac{3}{3+\alpha}$ denote the present density parameters for matter and the cosmological constant, respectively. Therefore, the model parameter $\sigma$ serves as a reliable indicator of the current model's deviation from the $\Lambda $CDM model, arising from the inclusion of the trace of the energy-momentum tensor terms.

As a dimensionless representation of the second order time derivative of the
scale factor of the universe, the deceleration parameter is defined as, $q=-%
\frac{\overset{..}{a}}{aH^{2}}$. Moreover, it can be expressed using the
Hubble parameter and its derivative in terms of the cosmological redshift as 
\cite{Xu},%
\begin{equation}
q\left( z\right) =-1+\left( 1+z\right) \frac{1}{H\left( z\right) }\frac{%
dH\left( z\right) }{dz}.  \label{22}
\end{equation}

The deceleration parameter $q\left( z\right) $ establishes the rate of
expansion of the universe. The value of $q\left( z\right) $ affects the
expansion of the universe. In other words, the universe exhibits
acceleration if $q\left( z\right) <0$ or deceleration if $q\left( z\right) >0
$ whereas $q\left( z\right) =0$ denotes expansion at a constant rate. For
the present parametrization of the effective EoS, the expression of the
deceleration parameter obtained by including Eq. (\ref{21}) into Eq. (\ref%
{22}) as,%
\begin{equation}
q\left( z\right) =-1+\frac{3 \alpha  (\sigma +8 \pi ) (z+1)^3}{3 \sigma  \left(\alpha  (z+1)^3+4\right)+16 \pi  \left(\alpha  (z+1)^3+3\right)}.
\label{qz}
\end{equation}

In addition, the third-order time derivative of the scale factor of the universe is represented without dimensions by the jerk parameter $j$, which
is defined as, $j=\frac{\overset{...}{a}}{aH^{3}}$. The deceleration
parameter and its derivative with respect to the cosmological redshift can
also be used to express the jerk parameter as,

\begin{equation}
j\left( z\right) =\left( 1+z\right) \frac{dq\left( z\right) }{dz}+q\left(
z\right) \left( 1+2q\left( z\right) \right) .
\end{equation}
\newline
According to popular belief, a cosmic jerk caused the transition from the
decelerating state to the accelerating state of the universe. For various
models with a positive value for the jerk parameter and a negative value for
the deceleration parameter, the universe transitions in this way \cite{Blandford, Sahni1}. For instance, the $\Lambda $CDM models have a fixed jerk
of $j\left( z\right) =1$. Moreover, the expression for the current model is

\begin{widetext}
\begin{equation}
j\left( z\right) =\frac{8 \left(9 \sigma ^2 \left(\alpha  (z+1)^3+2\right)+3 \pi  \sigma  \left(\alpha  (z+1)^3 \left(\alpha  (z+1)^3+28\right)+48\right)+32 \pi ^2 \left(\alpha  (z+1)^3+3\right)^2\right)}{\left(3 \sigma  \left(\alpha  (z+1)^3+4\right)+16 \pi  \left(\alpha  (z+1)^3+3\right)\right)^2}.
\end{equation}
\label{jz}
\end{widetext}

We are aware that certain physical parameters, including the EoS parameter,
the jerk parameter and the deceleration parameter, are crucial in the study
of the universe. An important new study in contemporary cosmology is on the
energy conditions derived from the Raychaudhuri equation \cite{Raychaudhuri}%
. Further, we have investigated the evolution of the most fundamental energy
conditions used in GR \cite{Arora1}. These conditions, place further
constraints on the validity of the proposed cosmological model, which are
specified as follows:%
\begin{widetext}
\begin{equation}
\text{WEC}:\rho =\frac{6 H_{0}^2 \left(\alpha  (z+1)^3+3\right) \left(\frac{3 \sigma  \left(\alpha  (z+1)^3+4\right)+16 \pi  \left(\alpha  (z+1)^3+3\right)}{3 (\alpha +4) \sigma +16 \pi  (\alpha +3)}\right)^{-\frac{\sigma }{3 \sigma +16 \pi }}}{3 (\alpha +4) \sigma +16 \pi  (\alpha +3)}\geq 0
\end{equation}%
\begin{equation}
\text{NEC}:\rho +p=\frac{6 \alpha  H_{0}^2 (z+1)^3 \left(\frac{3 \sigma  \left(\alpha  (z+1)^3+4\right)+16 \pi  \left(\alpha  (z+1)^3+3\right)}{3 (\alpha +4) \sigma +16 \pi  (\alpha +3)}\right)^{-\frac{\sigma }{3 \sigma +16 \pi }}}{3 (\alpha +4) \sigma +16 \pi  (\alpha +3)}%
\geq 0
\end{equation}%
\begin{equation}
\text{DEC}:\rho -p=\frac{6 H_{0}^2 \left(\alpha  (z+1)^3+6\right) \left(\frac{3 \sigma  \left(\alpha  (z+1)^3+4\right)+16 \pi  \left(\alpha  (z+1)^3+3\right)}{3 (\alpha +4) \sigma +16 \pi  (\alpha +3)}\right)^{-\frac{\sigma }{3 \sigma +16 \pi }}}{3 (\alpha +4) \sigma +16 \pi  (\alpha +3)}\geq 0
\end{equation}%
\begin{equation}
\text{SEC}:\rho +3p=\frac{6 H_{0}^2 \left(\alpha  (z+1)^3-6\right) \left(\frac{3 \sigma  \left(\alpha  (z+1)^3+4\right)+16 \pi  \left(\alpha  (z+1)^3+3\right)}{3 (\alpha +4) \sigma +16 \pi  (\alpha +3)}\right)^{-\frac{\sigma }{3 \sigma +16 \pi }}}{3 (\alpha +4) \sigma +16 \pi  (\alpha +3)}\geq 0
\end{equation}%
\end{widetext}where, WEC, NEC, DEC, and SEC represent weak energy
conditions, null energy conditions, dominant energy conditions, and strong
energy conditions, respectively. The null energy condition violation leads
to the remaining energy condition violation, which symbolizes the universe
expanding as its energy density is being depleted. Also, the strong energy
condition violation symbolizes the acceleration of the universe \cite{Visser}%
.

To study the evolution of the cosmological parameters of the model, we will
need to constrain the model's parameters i.e. $\alpha $ and $\sigma $ using
some recent observational data.

\section{Observational constraints from BAO, Hubble, and SNe datasets}

\label{sec4}

The estimate of the parameter values from the observational data is a
critical element of the parametrization method. The model has two
parameters: the parameter given by the $\omega _{eff}$ equation i.e. $\alpha 
$ and the coupling constant of model $\sigma $. In this part, we examine the
observational features of our present parametrization of the effective EoS.
To determine the model parameters $\alpha $ and $\sigma $ ranges that best
suit the data, we analyze six points of the baryonic acoustic oscillations
datasets (BAO), 57 points of the Hubble datasets, and 1048 points
from the Pantheon supernovae Ia sample datasets (SNe). We use the MCMC (Markov Chain Monte Carlo) approach in the \texttt{emcee} python library \cite%
{Mackey} together with the standard Bayesian method and likelihood function
to constrain the model parameters.

\subsection{BAO datasets}

The BAO distance datasets consists of BAO measurements at six different
redshifts for the 6dFGS, SDSS, and WiggleZ surveys. The comoving sound
horizon $r_{s}$ at the photon decoupling epoch $z_{\ast }$, which is defined
by the following relation, governs the characteristic scale of BAO:%
\begin{equation}
r_{s}(z_{\ast })=\frac{c}{\sqrt{3}}\int_{0}^{\frac{1}{1+z_{\ast }}}\frac{da}{%
a^{2}H(a)\sqrt{1+(3\Omega _{b0}/4\Omega _{\gamma 0})a}}.  \label{4b}
\end{equation}%
where $\Omega _{\gamma 0}$ represents the photon density parameter at the
moment and $\Omega _{b0}$ represents the baryon density parameter at the
moment.

In this work, to measure BAO, the following relationships are used,%
\begin{equation}
\triangle \theta =\frac{r_{s}}{d_{A}(z)},\text{ }d_{A}(z)=\int_{0}^{z}\frac{%
dz^{\prime }}{H(z^{\prime })},\text{ }\triangle z=H(z)r_{s}.  \label{4c}
\end{equation}

Here, $\triangle z$ is the observed redshift separation of the BAO
measurements in the two-point correlation function of the galaxy
distribution on the sky along the line of sight, $\triangle \theta $ denotes
the observed angular separation, and $d_{A}$ is the angular diameter
distance. Six-point BAO datasets for $d_{A}(z_{\ast })/D_{V}(z_{BAO})$
are obtained from the Refs. \cite{BAO1, BAO2, BAO3, BAO4, BAO5, BAO6}, where 
$d_{A}(z)$ denotes the co-moving angular diameter distance together with the
dilation scale $D_{V}(z)=\left[ d_{A}(z)^{2}z/H(z)\right] ^{1/3}$. Also, the
Planck measurements show that the redshift value for the photon decoupling
epoch is $z_{\ast }\approx 1091$. The chi-square function for the BAO sets
of data is defined as \cite{BAO6},

\begin{equation}  \label{4e}
\chi _{BAO}^{2}=X^{T}C^{-1}X\,,
\end{equation}

\begin{widetext}

where 

\begin{equation*}
X=\left( 
\begin{array}{c}
\frac{d_{A}(z_{\star })}{D_{V}(0.106)}-30.95 \\ 
\frac{d_{A}(z_{\star })}{D_{V}(0.2)}-17.55 \\ 
\frac{d_{A}(z_{\star })}{D_{V}(0.35)}-10.11 \\ 
\frac{d_{A}(z_{\star })}{D_{V}(0.44)}-8.44 \\ 
\frac{d_{A}(z_{\star })}{D_{V}(0.6)}-6.69 \\ 
\frac{d_{A}(z_{\star })}{D_{V}(0.73)}-5.45%
\end{array}%
\right) \,,
\end{equation*}

and,

\begin{equation*}
C^{-1}=\left( 
\begin{array}{cccccc}
0.48435 & -0.101383 & -0.164945 & -0.0305703 & -0.097874 & -0.106738 \\ 
-0.101383 & 3.2882 & -2.45497 & -0.0787898 & -0.252254 & -0.2751 \\ 
-0.164945 & -2.454987 & 9.55916 & -0.128187 & -0.410404 & -0.447574 \\ 
-0.0305703 & -0.0787898 & -0.128187 & 2.78728 & -2.75632 & 1.16437 \\ 
-0.097874 & -0.252254 & -0.410404 & -2.75632 & 14.9245 & -7.32441 \\ 
-0.106738 & -0.2751 & -0.447574 & 1.16437 & -7.32441 & 14.5022%
\end{array}%
\right) \,.
\end{equation*}

is the inverse of the covariance matrix \cite{BAO6}.

\end{widetext}

\begin{figure}[tbp]
\centering
\includegraphics[scale=0.75]{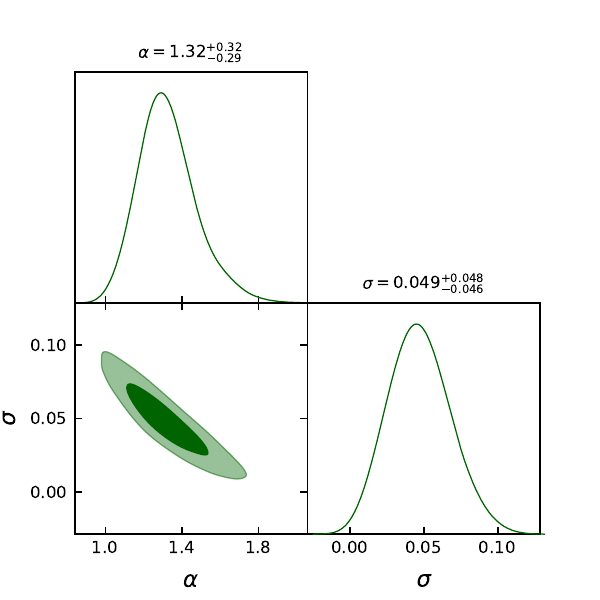}
\caption{The $1-\sigma$ and $2-\sigma$ likelihood contours are depicted in this figure, which represents the model parameter space constrained by the BAO datasets}
\label{ContourBAO}
\end{figure}

\begin{figure}[tbp]
\centering
\includegraphics[scale=0.75]{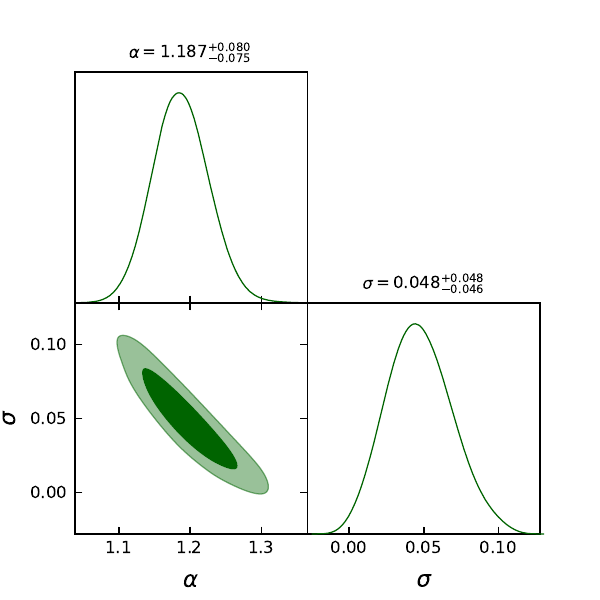}
\caption{The $1-\sigma$ and $2-\sigma$ likelihood contours are depicted in this figure, which represents the model parameter space constrained by the BAO+Hubble+SNe datasets.}
\label{H+SN+BAO}
\end{figure}

By minimizing the chi-square function for the BAO, we were able to determine
the parameter $\alpha $ and $\sigma $ ranges for the effective EoS model
that best suited the data. Fig. \ref{ContourBAO} presents the $1-\sigma $
and $2-\sigma $ likelihood contours for the model parameters $\alpha $ and $%
\sigma $ using the BAO datasets. The best-fit values that correspond to
observations are $\alpha =1.32_{-0.29}^{+0.32}$, and $\sigma
=0.049_{-0.046}^{+0.049}$.

\subsection{Hubble datasets}

Here, the Hubble parameter $H(z)$ has been measured by several teams. By
measuring the differential of cosmological redshift $z$ with regard to
cosmic time $t$, it is possible to estimate the value of $H(z)$ as $%
H(z)=-dz/[dt(1+z)]$. The model-independent value of the Hubble parameter may
be determined by measuring the quantity $dt$ since $dz$ is obtained from a
spectroscopic survey. We used an updated dataset of 57 Hubble measurements
in the $0.07\leq z\leq 2.41$ range of redshift {\cite{Sharov}}. The
differential age approach and line of sight BAO are two generally accepted
methods for determining the values of $H(z)$ at a particular redshift.
Further, we have considered $H_{0}=69$ $km/s/Mpc$ in our research \cite%
{Planck2020}. To calculate the mean values of the model parameters $\alpha $
and $\sigma $, the chi-square function is defined as,{\ }%
\begin{equation}
\chi _{Hubble}^{2}(\alpha ,\sigma )=\sum\limits_{i=1}^{57}\frac{%
[H_{th}(\alpha ,\sigma ,z_{i})-H_{obs}(z_{i})]^{2}}{\sigma _{H(z_{i})}^{2}},
\label{4a}
\end{equation}%
where, $H_{obs}$ is the observed value of the Hubble parameter, $H_{th}$ is
the predicted value of the Hubble parameter, and $\sigma _{H(z_{i})}$\ is
the standard error in the observed value of $H$.

\subsection{SNe datasets}

The most popular data sample for studying the late-time evolution of the
universe is data from SNe observations. 1048 type Ia supernovae with
redshifts $z$ between $0.01$ and $2.3$ make up the Pantheon samples, which
were assembled by Scolnic et al \cite{Scolnic}. It includes data from the
PanSTARSS1 Medium Deep Survey, SDSS, SNLS, and several low-z and HST
samples. The difference between the apparent magnitude ($m_{B}$) and
absolute magnitude ($M_{B}$) of the B band of the measured spectrum is the
distance modulus of type Ia SNe. It's described as,%
\begin{equation}
\mu (z)=5log_{10}d_{L}(z)+\mu _{0},  \label{4g}
\end{equation}%
where the luminosity distance, or $d_{L}(z)$, is defined in a spatially flat
FLRW universe as,%
\begin{equation}
d_{L}(z)=(1+z)\int_{0}^{z}\frac{cdz^{\prime }}{H(z^{\prime })},  \label{4f}
\end{equation}%
and%
\begin{equation}
\mu _{0}=5log(1/H_{0}Mpc)+25,
\end{equation}%
where $c$ represents the speed of light.

The theoretical distance modulus is correlated with the $\chi _{SNe}^{2}$
function for SNe as,%
\begin{equation}
\chi _{SNe}^{2}(\alpha ,\sigma )=\sum_{i=1}^{1048}\dfrac{\left[ \mu
_{obs}(z_{i})-\mu _{th}(\alpha ,\sigma ,z_{i})\right] ^{2}}{\sigma
^{2}(z_{i})},  \label{4i}
\end{equation}%
where $\mu _{obs}$ is the observed value, $\mu _{th}$ is the theoretical
value of the distance modulus, and $\sigma ^{2}(z_{i})$ is really the
standard error of the observed value.

\subsection{BAO+Hubble+SNe datasets}

In addition, we use the BAO, Hubble and SNe datasets to obtain joint
constraints for the parameters $\alpha $ and $\sigma $ using the total
likelihood function. From this point forward, the appropriate probability
and Chi-square functions are provided by,%
\begin{equation}
\mathcal{L}_{tot}=\mathcal{L}_{BAO}\times \mathcal{L}_{Hubble}\times 
\mathcal{L}_{SNe},
\end{equation}%
and%
\begin{equation}
\chi _{tot}^{2}=\chi _{BAO}^{2}+\chi _{Hubble}^{2}+\chi _{SNe}^{2}.
\end{equation}

Fig. \ref{H+SN+BAO} depicts the $1-\sigma $ and $2-\sigma $ likelihood
contours for the model parameters $\alpha $ and $\sigma $ using the
BAO+Hubble+SNe datasets. The best-fit values that correspond to
observations are $\alpha =1.187_{-0.075}^{+0.080}$, and $\sigma
=0.048_{-0.046}^{+0.048}$.

Fig. \ref{ErrorHubble} compared our model with the well-known $\Lambda $CDM
cosmology model while taking into account the value matter density parameter 
$\Omega _{m_{0}}=0.315$ \cite{Planck2020}. The Hubble dataset findings, which
total 57 data points and their associated error, are also shown in the
figure, allowing for an easy comparison of the two models.

A comparison between our $f(Q,T)$ model and the well-known $\Lambda $CDM model in
cosmology is shown in Fig. \ref{ErrorSNe}. The Pantheon experimental
findings, 1048 data points, and their error are also shown in the graphic,
allowing for a direct comparison between the two models

\begin{widetext}

\begin{figure}[H]
\centerline{\includegraphics[width=18cm,height=6cm]{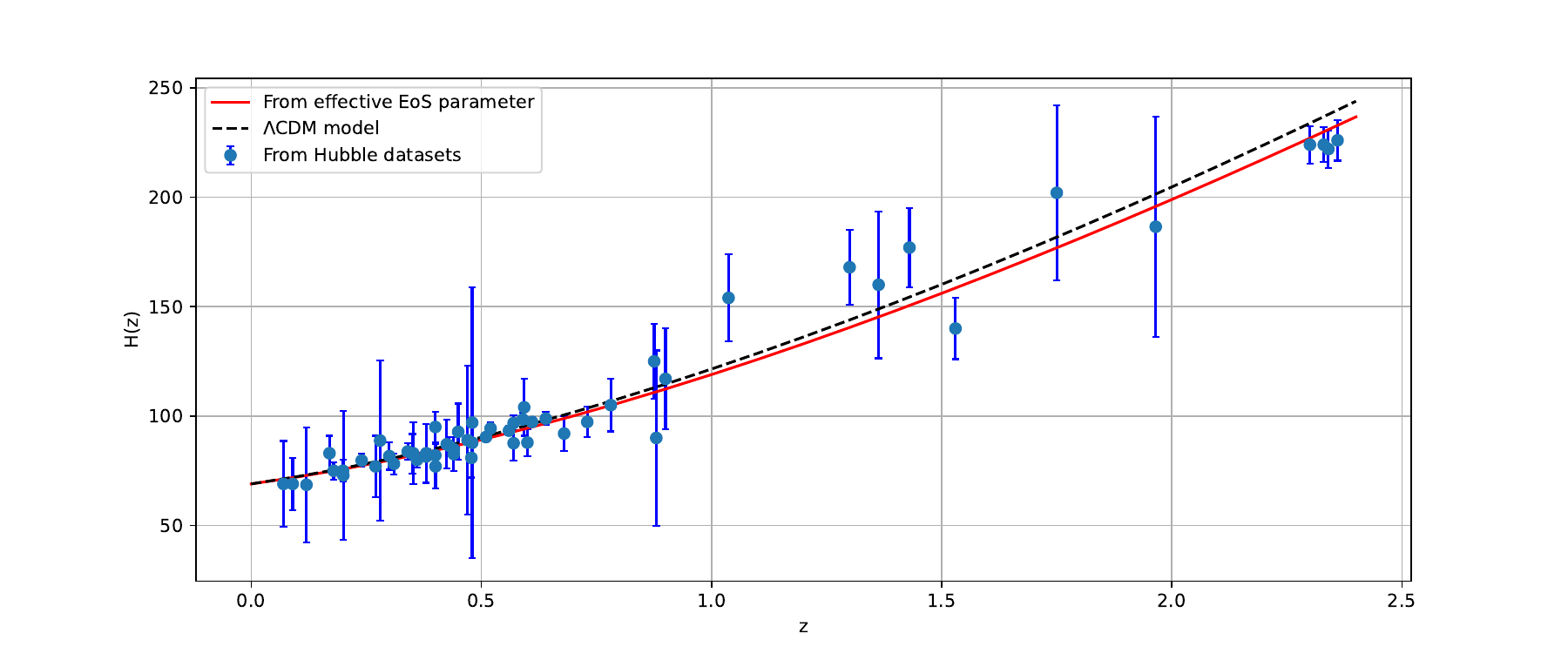}}
\caption{The graph illustrates a satisfactory correspondence between our $f(Q,T)$ theory and the 57 data points from the Hubble datasets, as depicted by the red curve (in units of $km/s/Mpc$), in comparison to the black dashed line representing $\Lambda$CDM, in a plot of $H(z)$ against the redshift $z$.}
\label{ErrorHubble}
\end{figure}

\begin{figure}[H]
\centerline{\includegraphics[width=18cm,height=6cm]{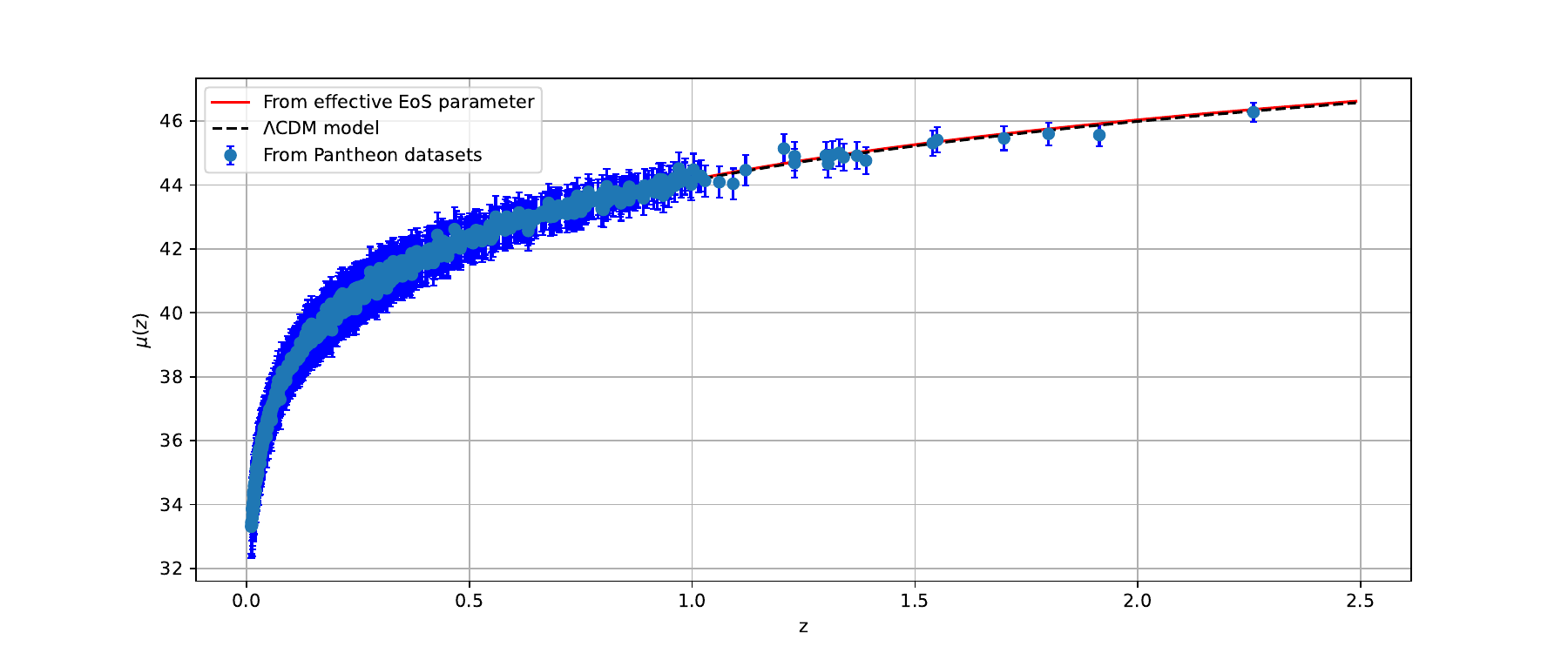}}
\caption{The graph illustrates a satisfactory correspondence between our $f(Q,T)$ theory and the 1048 points from the Pantheon datasets, as depicted by the red curve, in comparison to the black dashed line representing $\Lambda$CDM, in a plot of $\mu(z)$ against the redshift $z$.}
\label{ErrorSNe}
\end{figure}

\end{widetext}

\begin{figure}[]
\includegraphics[width=8.5 cm]{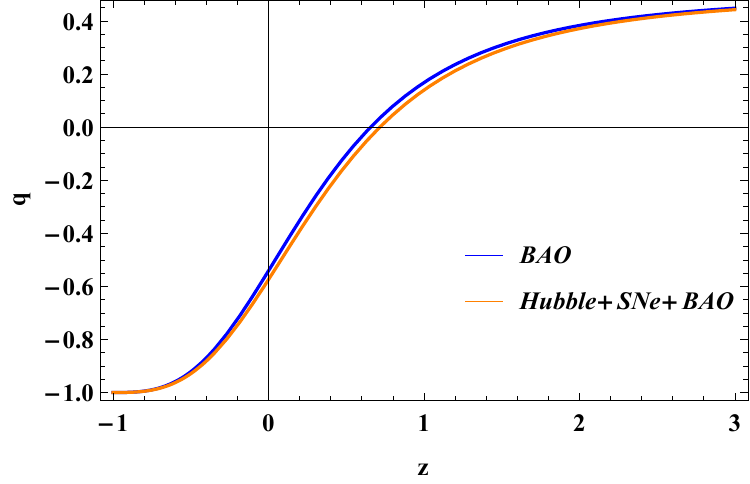}
\caption{The graph above shows the relationship between the deceleration
parameter ($q$) and redshift ($z$).}
\label{Fig-qz}
\end{figure}

\begin{figure}[tbp]
\includegraphics[width=8.5 cm]{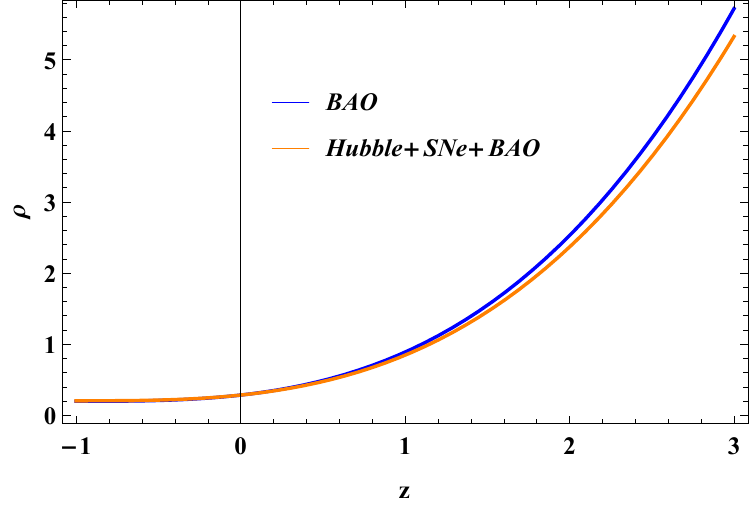}
\caption{The graph above shows the relationship between the energy density ($%
\protect\rho $) and redshift ($z$).}
\label{Fig-rho}
\end{figure}

\begin{figure}[tbp]
\includegraphics[width=8.5 cm]{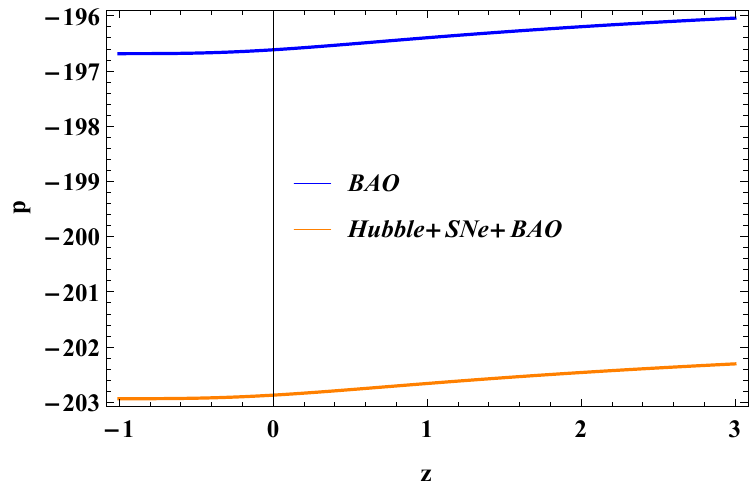}
\caption{The graph above shows the relationship between the pressure ($p$)
and redshift ($z$).}
\label{Fig-p}
\end{figure}

\begin{figure}[tbp]
\includegraphics[width=8.5 cm]{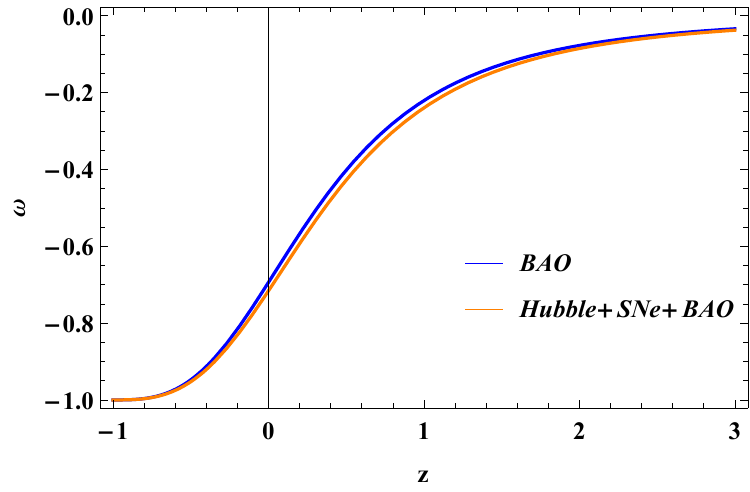}
\caption{The graph above shows the relationship between the effective EoS
parameter ($\protect\omega $) and redshift ($z$).}
\label{Fig-EoS}
\end{figure}

\begin{figure}[]
\centering\includegraphics[width=8.5 cm]{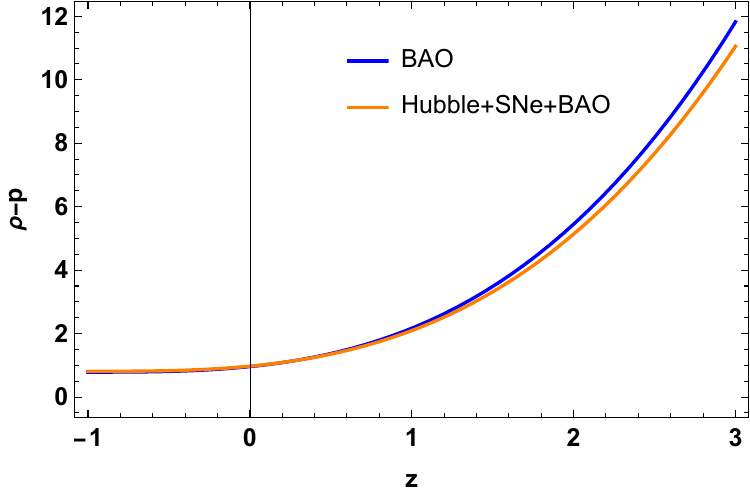}
\caption{The graph above shows the relationship between the DEC condition
and redshift ($z$).}
\label{Fig-DEC}
\end{figure}

\begin{figure}[]
\includegraphics[width=8.5 cm]{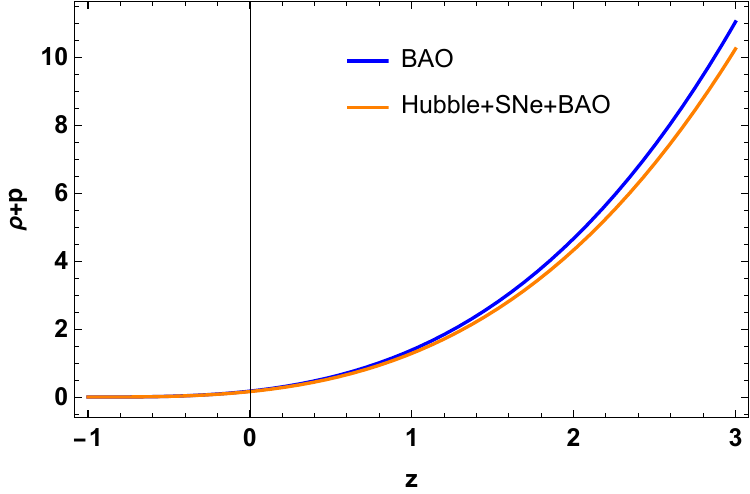}
\caption{The graph above shows the relationship between the NEC condition
and redshift ($z$).}
\label{Fig-NEC}
\end{figure}

\begin{figure}[]
\includegraphics[width=8.5 cm]{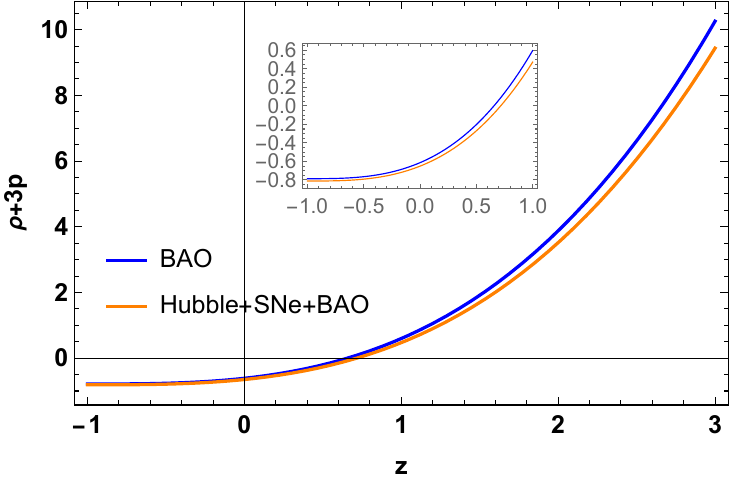}
\caption{The graph above shows the relationship between the SEC condition
and redshift ($z$).}
\label{Fig-SEC}
\end{figure}

\begin{figure}[]
\includegraphics[width=8.7 cm]{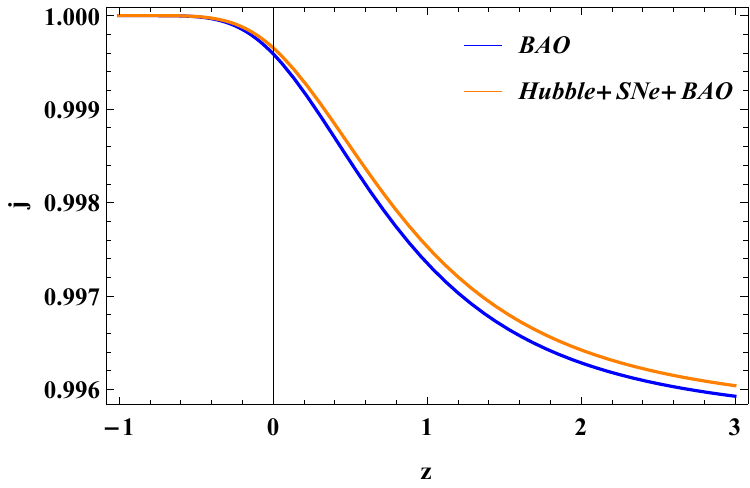}
\caption{The graph above shows the relationship between the jerk parameter ($%
j$) and redshift ($z$).}
\label{Fig-jerk}
\end{figure}

\section{Discussion and conclusion}
\label{sec5}

The parametrization approach has been extensively examined in the literature as a method to address various cosmological challenges. These challenges include the initial singularity problem, the issue of all-time decelerating expansion, the horizon problem, Hubble tension, and more \cite{Pacif1,Pacif2}. In our investigation, we introduce a parametric representation for the effective EoS parameter $\omega_{eff}$, as a fundamental aspect of constructing the model within the framework of $f\left( Q,T\right)$ gravity theory. This theoretical framework is based on the interplay between the energy-momentum trace $T$ and the non-metricity scalar $Q$. The use of a parametric form for $\omega_{eff}$ allows for a more flexible and nuanced exploration of the dynamic relationship between these key variables, providing a comprehensive understanding of their interdependence within the context of $f\left( Q,T\right)$ gravity theory.

We specifically examine the functional form $f(Q,T)=-Q+\sigma T$, where $\sigma $ is a free parameter. To solve the field equations for $H\left( z\right) $, we employ a parametric form of the $\omega {eff}$ parameter as a function of redshift $z$. Furthermore, by utilizing BAO datasets and combined BAO+Hubble+SNe datasets, we determine the best-fit values of the model parameters. The results of the best fit are $\alpha =1.32_{-0.29}^{+0.32}$, and $\sigma
=0.049_{-0.046}^{+0.049}$ for the BAO datasets and $\alpha
=1.187_{-0.075}^{+0.080}$, and $\sigma =0.048_{-0.046}^{+0.048}$ for the
BAO+Hubble+SNe datasets. In addition, we investigate the behavior of the deceleration parameter, energy density, pressure, effective EoS parameter, and jerk parameter for the constrained values of the model parameters.

In any cosmological model, the deceleration parameter $q$ is a crucial parameter used to characterize the periods of decelerated expansion ($q>0$) and accelerated expansion ($q<0$) of the universe. As shown in Eq. (\ref{qz}), it is evident that $q$ approaches -1 as the redshift $z$ approaches -1. The relationship between the deceleration parameter $q$ and the redshift $z$ is illustrated in Fig. \ref{Fig-qz}. To further investigate the nature of the cosmological parameters, we have incorporated the values of the pair ($\alpha ,\sigma $) obtained from observational data into all our analyses and graphs. It is observed
that $q$ exhibits a transition from early positive values to current
negative ones. In this way, the constructed model of the universe develops
from an early decelerated period to the present accelerated period. With the
observational data, this characteristic is in strong accord. According to
the values of the model parameters constrained by BAO, and the combined
BAO+Hubble+SNe datasets, the transition redshift are $z_{tr}=0.6589$ and 
$z_{tr}=0.7259$, respectively \cite{z1,z2}. Furthermore, the current value of the
deceleration parameter is $q_{0}=-0.534$ for the BAO datasets and $%
q_{0}=-0.581$ for the combined BAO+Hubble+SNe datasets \cite{q1,q2}.

The evolution of energy density and pressure, respectively, versus redshift $%
z$ is depicted in Figs. \ref{Fig-rho}, and \ref{Fig-p}. It has been shown
that the energy density increases positively with redshift $z$. In addition,
from the early era to the present, pressure p has negative values. The
current acceleration of the universe may be caused by the negative cosmic
pressure. We also examined the behavior of the effective EoS parameter,
which is depicted in Fig. \ref{Fig-EoS}. The relationship between the
previous energy density and pressure is established by the EoS parameter as
a whole i.e. $\omega =\frac{p}{\rho }$. The matter phase at $\omega =0$ is
one of the typical phases that may be identified using the EoS parameter.
The radiation-dominated phase is then shown by $\omega =\frac{1}{3}$, while $%
\omega =-1$ represents the $\Lambda $CDM model. In addition, the recently
discussed accelerating period of the universe, which contains the
quintessence ($-1<\omega <-\frac{1}{3}$) and phantom era ($\omega <-1$), is
shown when $\omega <-\frac{1}{3}$. According to the constrained values of $%
\alpha $ and $\sigma $, the effective EoS parameter in Fig. \ref{Fig-EoS}
exhibits behavior like quintessence. Also, the value of effective EoS
parameter at $z=0$ is $\omega _{0}=-0.7106$  for the BAO datasets and $%
\omega _{0}=-0.7268$ for the combined BAO+Hubble+SNe datasets, which is
a definite indication of an accelerating phase \cite{O1,O2,O3}.

The primary objective of energy conditions in cosmology is to assess the expansion behavior of the universe. These conditions serve as crucial indicators of the energy-momentum content of the universe and its impact on cosmic dynamics. Specifically, the violation of the NEC ($\rho+p \geq0$) implies a violation of the weak ($\rho \geq0$), dominant ($\rho-p \geq0$), and strong energy conditions ($\rho+3p \geq0$). This violation indicates a decrease in energy density as the universe expands, reflecting the phenomenon of energy dissipation or dilution over cosmic time. Moreover, the violation of the SEC ($\rho+3p \geq0$) is particularly significant as it signifies the acceleration of the universe. In our analysis of the energy conditions for the physical model, as shown in Figs. \ref{Fig-DEC}-\ref{Fig-SEC}, we have observed a consistent fulfillment of the WEC ($\rho \geq0$), NEC ($\rho+p \geq0$), and DEC ($\rho-p \geq0$) throughout the cosmic evolution. These conditions play a crucial role in characterizing the energy-momentum content of the universe and are indicative of its stability and behavior. Notably, the SEC ($\rho+3p \geq0$) is found to be violated at late cosmic times, suggesting the presence of exotic forms of energy or modifications to the gravitational theory in these regimes. This violation opens up intriguing possibilities for understanding the nature of DE and the dynamics of the universe at large scales. The relationship between the jerk parameter $j$ and the redshift $z$ is seen
in Fig. \ref{Fig-jerk}. The graphic shows that throughout cosmic history, $j(z)$ has remained positive. According to Sahni et al. \cite{Sahni1}%
, the jerk parameter for the $\Lambda $CDM model has a value of $j\left(
z\right) =1$. According to Eq. (\ref{jz}), $j\left( z\right) \neq 1$ at
currently, meaning that the model is currently deviating from $\Lambda $CDM.
The resultant model, however, behaves in the future as a standard $\Lambda $%
CDM, i.e. $j(z)\rightarrow 1$ as $z\rightarrow -1$. With current observational data, the
results of the model are in strong agreement. The established model is
regarded as a helpful reference point for the analysis of FLRW models that
include a parametric form of the effective EoS parameter in
the $f(Q,T)$ theory of gravity.

\section*{Acknowledgments}
AE thanks the National Research Foundation of South Africa for the award of a postdoctoral fellowship.



\begin{thebibliography}{99}
\bibitem{Riess} A.G. Riess et al., \textit{Astron. J.} \textbf{116}, 1009
(1998).

\bibitem{Perlmutter} S. Perlmutter et al., \textit{Astrophys. J.} \textbf{517%
}, 565 (1999).

\bibitem{T.Koivisto} T. Koivisto, D.F. Mota, \textit{Phys. Rev. D} \textbf{73%
}, 083502 (2006).

\bibitem{S.F.} S.F. Daniel, \textit{Phys. Rev. D} \textbf{77}, 103513 (2008).

\bibitem{D.J.} D.J. Eisenstein et al., \textit{Astrophys. J.} \textbf{633},
560 (2005).

\bibitem{W.J.} W.J. Percival at el., \textit{Mon. Not. R. Astron. Soc.} 
\textbf{401}, 2148 (2010).

\bibitem{R.R.} R.R. Caldwell, M. Doran, \textit{Phys. Rev. D} \textbf{69},
103517 (2004).

\bibitem{Z.Y.} Z.Y. Huang et al., \textit{J. Cosm. Astrop. Phys.} \textbf{%
0605}, 013 (2006).

\bibitem{weinberg/1989} S. Weinberg, \textit{Rev. Mod. Phys.} \textbf{61}, 1
(1989).

\bibitem{Steinhardt/1999} P.J. Steinhardt et al., \textit{Phys. Rev. Lett.}, \textbf{59}, 123504 (1999).

\bibitem{Hinshaw} G. Hinshaw et al., \textit{Astrophys. J. Suppl.} 
\textbf{208}, 19 (2013).

\bibitem{Planck2020} N. Aghanim et al., \textit{Astron. Astrophys.} 
\textbf{641}, A6 (2020).

\bibitem{Kamenshchik} A.Y. Kamenshchik et al., \textit{Phys. Lett. B} 
\textbf{511}, 265 (2001).

\bibitem{Sahni} V. Sahni and Y. Shtanov, \textit{J. Cosmol. Astropart. Phys.}
\textbf{11}, 014 (2003).

\bibitem{Li} M. Li, \textit{Phys. Lett. B} \textbf{603}, 1 (2004).

\bibitem{Cai} R. G. Cai, \textit{Phys. Lett. B} \textbf{657}, 228 (2007).

\bibitem{Padmanabhan} T. Padmanabhan, \textit{Phys. Rev. D} \textbf{66},
02131 (2002).

\bibitem{Caldwell} R. R. Caldwell, \textit{Phys. Lett. B} \textbf{545}, 23
(2002).

\bibitem{Nojiri} S. Nojiri and S. D. Odintsov, \textit{Phys. Lett. B} 
\textbf{562}, 147 (2003).

\bibitem{Copeland} E. J. Copeland et al., \textit{Int. J. Mod. Phys. D} 
\textbf{15}, 1753 (2006).

\bibitem{Harko} T. Harko et al., \textit{Phys. Rev. D} \textbf{84}, 024020
(2011).

\bibitem{Nojiri1} S. Nojiri et al., \textit{Phys. Rep.} \textbf{692}, 1
(2017).

\bibitem{Harko/2010} T. Harko and F. S. N. Lobo, \textit{Eur. Phys. J. C} \textbf{70}, 373-379 (2010).

\bibitem{THK-2} T. Harko, \textit{Phys. Rev. D} \textbf{81}, 084050 (2010).

\bibitem{THK-3} T. Harko, \textit{Phys. Rev. D} \textbf{81}, 044021 (2010).
 
\bibitem{V.F.-2} V. Faraoni, \textit{Phys. Rev. D} \textbf{76}, 127501 (2007).

\bibitem{fG1} A. De Felice and S. Tsujikawa, \textit{Phys. Lett. B} \textbf{675}, 1-8 (2009).

\bibitem{Laurentis/2015} M. De Laurentis, M. Paolella, S. Capozziello, \textit{Phys. Rev. D}, \textbf{91}, 083531 (2015).

\bibitem{Gomez/2012} A. de la Cruz-Dombriz, D. S-Gomez, \textit{Class. Quantum Grav.},  \textbf{29} 245014 (2012). 

\bibitem{Riemann} B. Riemann, Habilitationsschrift, 1854, \textit{Abh. Koniglichen Ges. Wiss. Gott.} \textbf{13}, 1 (1867).

\bibitem{Weyl} H. Weyl, \textit{Sitz. Preuss. Akad. Wiss}. \textbf{465}, 1
(1918).

\bibitem{Moller} C. Moller, \textit{Mat. Fys. Skr. Dan. Vid. Selsk}. \textbf{%
1}, 10 (1961).

\bibitem{Pellegrini} C. Pellegrini and J. Plebanski, \textit{Mat. Fys. Skr.
Dan. Vid. Selsk.} \textbf{2}, 4 (1963).

\bibitem{Hayashi} K. Hayashi, T. Shirafuji, \textit{Phys. Rev. D} \textbf{19}%
, 3524 (1979).

\bibitem{Jimenez1} J. B. Jimenez, L. Heisenberg and T. Koivisto, \textit{%
Phys. Rev. D} \textbf{98}, 044048 (2018).

\bibitem{Jimenez2} J. B. Jimenez et al., \textit{Phys. Rev. D} \textbf{101},
10 (2020).

\bibitem{Lazkoz} R. Lazkoz, F.S.N. Lobo, M. Ortiz-Bano, V. Salzano, \textit{%
Phys. Rev. D} \textbf{100}, 104027 (2019).

\bibitem{Frusciante} N. Frusciante, \textit{Phys. Rev. D} \textbf{103}, 4
(2021).

\bibitem{MK1} M. Koussour et al., \textit{Phys. Dark universe} \textbf{36},
101051 (2022).

\bibitem{MK2} M. Koussour and M. Bennai \textit{Chin. J. Phys. } \textbf{379},
339-347 (2022).

\bibitem{MK3} M. Koussour et al. \textit{Phys. Ann. Phys.} \textbf{445},
169092 (2022).

\bibitem{MK4} M. Koussour and A. De, \textit{Eur. Phys. J. C} \textbf{83}, 400 (2023).

\bibitem{MK5} M. Koussour et al., \textit{Fortschr. Phys.} \textbf{71}, 2200172 (2023).

\bibitem{MK6} M. Koussour et al., \textit{Nucl. Phys. B} \textbf{990}, 116158 (2023).

\bibitem{MK7} M. Koussour et al., \textit{J. High Energy Phys.} \textbf{37}, 15-24 (2023).

\bibitem{MK8} M. Koussour et al., \textit{J. High Energy Astrophys, } \textbf{35}, 43-51 (2022).

\bibitem{Harko1} T. Harko et al., \textit{Phys. Rev. D} \textbf{98}, 8
(2018).

\bibitem{Xu} Y. Xu et al., \textit{Eur. Phys. J. C} \textbf{79,} 708 (2019).

\bibitem{Pati} L. Pati, B. Mishra, and S. K. Tripathy, \textit{Phys. Scr.} 
\textbf{96}, 105003 (2021).

\bibitem{Ag} A.S. Agrawal et al., \textit{Phys. Dark Univ. }\textbf{33},
100863 (2021).

\bibitem{Bhattacharjee} S. Bhattacharjee, P.K. Sahoo, \textit{Eur. Phys. J. C%
} \textbf{80}, 289 (2020).

\bibitem{Pacif1} S. K. J. Pacif, \textit{Eur. Phys. J. Plus}, \textbf{135},
10 (2020).

\bibitem{Pacif2} S. K. J. Pacif, R. Myrzakulov and S. Myrzakul, \textit{Int.
J. Geom. Methods Mod.}, \textbf{14}, 07, (2017).

\bibitem{Arora/2020} S. Arora et al., \textit{Phys. Dark Univ.} \textbf{30}, 100664 (2020).

\bibitem{Loo/2023} Tee-How Loo, M. Koussour, and Avik De, \textit{Ann. Phys.} \textbf{454}, 169333 (2023).

\bibitem{Tayde/2022} M. Tayde, Z. Hassan, P. K. Sahoo, \textit{Chinese Phys. C} \textbf{46}, 115101 (2022).

\bibitem{CPL1} M. Chevallier and D. Polarski, \textit{Int. J. Mod. Phys. D} \textbf{10}, 213 (2001).

\bibitem{CPL2} E. V. Linder, \textit{Phys. Rev. Lett.} \textbf{90}, 091301 (2003).

\bibitem{JBP} H. K. Jassal, J. S. Bagla, T. Padmanabhan, \textit{Mon. Not. R. Astron. Soc. Lett.} \textbf{356}, L11-L16 (2005).

\bibitem{MZ} J.-Z. Ma and X. Zhang, \textit{Phys. Lett. B} \textbf{669}, 233 (2011).

\bibitem{Mukherjee} A. Mukherjee and N. Banerjee, \textit{Eur. Phys. J. Plus}%
, \textbf{130}, 10 (2015).

\bibitem{Gong/2005} Y. Gong and Y. Zhang, \textit{Phys. Rev. D}, \textbf{72}, 043518 (2005).

\bibitem{Yang/2019} W. Yang et al., \textit{Phys. Rev. D}, \textbf{99}, 043543 (2019).

\bibitem{Blandford} R. D. Blandford et al., \textit{Observing Dark Energy}, 
\textbf{339}, 27 (2005) arXiv:astro-ph/0408279.

\bibitem{Sahni1} V. Sahni, T. D. Saini, A. A. Starobinsky, U. Alam, \textit{%
JETP Lett.} \textbf{77}, 201 (2003).

\bibitem{Raychaudhuri} A. Raychaudhuri, \textit{Phys. Rev.} \textbf{98},
1123 (1955).

\bibitem{Arora1} S. Arora and P. K. Sahoo, \textit{Phys. Scr.} \textbf{95},
9 (2020).

\bibitem{Visser} M. Visser, C Barcelo, \textit{COSMO-99}, 98 (2000).

\bibitem{Mackey} D. F. Mackey et al., \textit{Publ. Astron. Soc. Pac.} 
\textbf{125}, 306 (2013).

\bibitem{BAO1} C. Blake et al., \textit{\ Mon. Not. Roy. Astron. Soc.} 
\textbf{418}, 1707 (2011).

\bibitem{BAO2} W. J. Percival et al., \textit{Mon. Not. Roy. Astron. Soc.} 
\textbf{401}, 2148 (2010).

\bibitem{BAO3} F. Beutler et al., \textit{\ Mon. Not. Roy. Astron. Soc.} 
\textbf{416}, 3017 (2011).

\bibitem{BAO4} N. Jarosik et al., \textit{\ Astrophys. J. Suppl.} \textbf{192%
}, 14 (2011).

\bibitem{BAO5} D. J. Eisenstein et al., \textit{\ Astrophys. J.} \textbf{633}%
, 560 (2005).

\bibitem{BAO6} R. Giostri et al.,\textit{\ J. Cosm. Astropart. Phys.} 
\textbf{1203}, 027 (2012).

\bibitem{Sharov} G.S. Sharov, V.O. Vasilie, \textit{Mathematical Modelling
and Geometry} \textbf{6}, 1 (2018).

\bibitem{Scolnic} D.M. Scolnic et al., \textit{ApJ} \textbf{859}, 101(2018).

\bibitem{z1} J. Roman-Garza et al.,\textit{Eur. Phys. J. C} 
\textbf{79}, 890 (2019).

\bibitem{z2} J.F. Jesus et al.,\textit{J. Cosmol. Astropart. Phys.} 
\textbf{04}, 053 (2020).

\bibitem{q1} A. Hernandez-Almada et al.,\textit{Eur. Phys. J. C} 
\textbf{79}, 12 (2019).

\bibitem{q2} S. Basilakos, F. Bauer, and J. Sola,\textit{J. Cosmol. Astropart. Phys.} 
\textbf{01}, 050 (2012).

\bibitem{O1} C. Gruber and O. Luongo, \textit{Phys. Rev. D} \textbf{89}, 103506 (2014).

\bibitem{O2} S. Arora, A. Parida and P.K. Sahoo , \textit{Eur. Phys. J. C} \textbf{81},  1-7 (2021). 

\bibitem{O3} N. Myrzakulov et al., \textit{Eur. Phys. J. Plus} \textbf{138}, 852 (2023). 

\end{thebibliography}
\end{document}